\documentclass[a4paper,10pt]{article}
\usepackage{graphicx}
\usepackage{comment}
\usepackage{amssymb}
\usepackage{amsmath}
\usepackage{nicefrac}
\usepackage{dcolumn}
\usepackage{bm}
\usepackage{comment}
\usepackage{multirow}
\usepackage{cite}
\usepackage{xcolor}
\usepackage{url}
\usepackage{mathrsfs}
\usepackage{supertabular}
\usepackage[normalem]{ulem}
\usepackage{lscape}
\usepackage{soul}
\usepackage{subcaption} 
\usepackage{bold-extra}
\usepackage{hyperref}
\usepackage{dcolumn,ulem,enumitem}
\usepackage{bm}
\usepackage{xcolor}
\usepackage{tikz}

\begin{document}
            
\newcommand{\bin}[2]{\left(\begin{array}{c}\!#1\!\\\!#2\!\end{array}\right)}
\newcommand{\threej}[6]{\left(\begin{array}{ccc}#1 & #2 & #3 \\ #4 & #5 & #6 \end{array}\right)}
\newcommand{\sixj}[6]{\left\{\begin{array}{ccc}#1 & #2 & #3 \\ #4 & #5 & #6 \end{array}\right\}}
\newcommand{\regge}[9]{\left[\begin{array}{ccc}#1 & #2 & #3 \\ #4 & #5 & #6 \\ #7 & #8 & #9 \end{array}\right]}
\newcommand{\La}[6]{\left[\begin{array}{ccc}#1 & #2 & #3 \\ #4 & #5 & #6 \end{array}\right]}
\newcommand{\hj}{\hat{J}}
\newcommand{\hux}{\hat{J}_{1x}}
\newcommand{\hdx}{\hat{J}_{2x}}
\newcommand{\huy}{\hat{J}_{1y}}
\newcommand{\hdy}{\hat{J}_{2y}}
\newcommand{\huz}{\hat{J}_{1z}}
\newcommand{\hdz}{\hat{J}_{2z}}
\newcommand{\hup}{\hat{J}_1^+}
\newcommand{\hum}{\hat{J}_1^-}
\newcommand{\hdp}{\hat{J}_2^+}
\newcommand{\hdm}{\hat{J}_2^-}

\huge

\begin{center}
Excitation and ionization by electron impact in transition and super-transition arrays
\end{center}

\vspace{0.5cm}

\large

\begin{center}
Djamel Benredjem$^{a,}$\footnote{djamel.benredjem@universite-paris-saclay.fr} and Jean-Christophe Pain$^{b,c}$
\end{center}

\normalsize

\begin{center}
\it $^a$Laboratoire Aim\'e Cotton, Universit\'e Paris-Saclay, Orsay, France\\
\it $^b$CEA, DAM, DIF, F-91297 Arpajon, France\\
\it $^c$Universit\'e Paris-Saclay, CEA, Laboratoire Mati\`ere en Conditions Extr\^emes,\\
\it 91680 Bruy\`eres-le-Ch\^atel, France\\
\end{center}

\vspace{0.5cm}

\begin{abstract}
This study investigates the ionization and excitation processes induced by electron impact between two configurations or superconfigurations. Rate coefficients are calculated for transition arrays or super-transition arrays rather than level-to-level transitions. Special attention is given to a series of oxygen-like ions relevant to inertial confinement fusion, specifically silicon, germanium, argon, and krypton. Calculations are performed over a wide temperature range, from 50 to 3000 eV. To facilitate the determination of rates at any temperature and ion charge, the computed rates are fitted with a two-dimensional Chebyshev polynomial expansion, yielding a set of coefficients for practical applications. Additionally, an extension of the Clenshaw algorithm to two dimensions, using the Chebyshev coefficients, is proposed to address numerical challenges and enhance computational efficiency.
\end{abstract}

\section{\label{sec:Introduction}Introduction}
The knowledge of the rates of microscopic processes and their dependence on temperature and ion charge is essential for the evolution of hot and dense plasmas, especially in the context of Inertial Confinement Fusion (ICF) \cite{Atzeni2004,Abu2024}. Among these processes, electron-impact excitation and ionization are particularly significant and have been extensively studied both experimentally and theoretically. In ICF, dopants are embedded into the fuel capsule or ablator to optimize implosion dynamics, enhance energy absorption, and facilitate plasma diagnostics. These dopants are selected based on their atomic properties, such as opacity, ionization potential, and radiative characteristics, which affect implosion efficiency and stability. Dopants like silicon, germanium \cite{Hu2012,Hill2012,Benredjem2015,Li2015}, and bromine can be added to the ablator to control X-ray absorption and reduce hydrodynamic instabilities. Their high opacity ensures effective absorption and re-emission of X-rays from the Hohlraum, promoting a uniform drive and minimizing asymmetries in the implosion. Additionally, they mitigate preheat effects by shielding the fuel from early energy deposition, maintaining the required compression. By smoothing the ablation front, they also help suppressing Rayleigh-Taylor instabilities. To improve the stability of the fuel-capsule interface and refine the imploding shell profile, capsules may be infused with a small fraction of high-$Z$ material, such as in indirect-drive experiments with tungsten-doped high-density carbon capsules \cite{Hopkins2018,Li2021}.\\
\indent Mid- and low-Z dopants, like krypton, argon, silicon and boron serve specific roles based on their placement in the capsule. Rare gases like argon, krypton and xenon, added in trace amounts to the deuterium-tritium (DT) fuel, act as spectroscopic tracers, enabling precise measurements of temperature and density via X-ray emission. Alternative fuel mixtures, like $_3$He-doped deuterium, are used to tailor neutron production and study fusion reaction kinetics.\\
\indent Understanding the rates of various processes is essential for modeling the ionic populations in non-LTE (non Local Thermodynamic Equilibrium) hot plasmas \cite{Colombant1973,Ralchenko2016}, which are described by a collisional-radiative system of equations (see for instance Ref. \cite{Florido2009}). In this study, we focus on the rate variation in O-like ions Si$^{6+}$, Ar$^{10+}$, Ti$^{14+}$, Fe$^{18+}$, Cu$^{21+}$, Ge$^{24+}$ and Kr$^{28+}$. Specifically, we examine transitions between two configurations or superconfigurations, referred respectively to as a transition array (ensemble of transitions between two configurations) \cite{Bauche1988} or super-transition arrays (STA, ensemble of transitions between two superconfigurations) \cite{Peyrusse1999}. The cross sections and rates for these arrays encompass multiple level-to-level transitions. In this work, we use the Flexible Atomic Code (FAC) \cite{Gu2008} to calculate electron impact ionization (EII) \cite{Fontes1993} and electron-impact excitation (EIE) cross sections, with level-to-level values computed using the relativistic Distorted-Wave (DW) method. This approach extends the factorization-interpolation procedures developed for EIE cross sections and allows for the inclusion of the most general configuration mixing. Other methods, such as Coulomb-Born-Exchange (CBE) or Binary-Encounter-Dipole (BED), can also be implemented. An investigation of processes involved in hot and dense media needs to take into account temperature variations in calculated cross sections and rates. Because fine-structure calculations introduce additional complexity on top of radiative-collisional calculations, many computational codes employ statistical approaches that cluster levels into configurations \cite{Ralchenko2016,Hansen2023} or even merge configurations into superconfigurations \cite{Barshalom1997,Peyrusse1999,Peyrusse2000}. This strategy requires defining rates among these aggregated entities. In addition, it is important to evaluate how this averaging process affects the results \cite{Oreg1997,Poirier2007}.\\
\indent In Section \ref{sec:Rates}, we define the cross sections and rates for a transition array and an STA. The population fractions of the relativistic energy levels are assumed to follow a Maxwellian distribution, implying that both the cross sections and rates depend on temperature. In Section III, we show that the calculated rate can be accurately fitted using two-dimensional Chebyshev polynomial expansions \cite{Rivlin1974,Oliver1977,Wetterling1991,Faussurier2017}. The fits are highly satisfactory and yield useful coefficients for the Chebyshev expansion in two dimensions (temperature and ion charge). In Section IV, we present numerical results for excitation and direct ionization in transition arrays and STAs, showing how the rates vary with temperature and ion charge. In Section V, we present an alternative to the polynomial expansions by generalizing the Clenshaw algorithm \cite{Clenshaw1957} to two dimensions. This provides a compact representation that accelerates the calculation while avoiding numerical difficulties. The Chebyshev expansion and the generalized Clenshaw algorithm yield identical results.

\section{\label{sec:Rates}Rates in transition arrays and super-transition arrays}
The cross section for excitation or ionization between an initial (\textit{resp.} final) level $i$ (\textit{resp.} $f$) is named $\sigma_{if}(E)$, where $E$ is the energy of the projectile (electron) and $E_{if}$ the transition energy. The corresponding rate coefficient is defined as
\begin{equation*}
    q_{if}=\int_{E_{if}}^{\infty} \sigma_{if}(E)\,v\,\rho(E)\,dE,
\end{equation*}
where $v=\sqrt{2E/m_e}$ is the velocity of the projectile, with $m_e$ designating its mass, and $\rho$ the normalized distribution of the free electrons at temperature $T_e$. Assuming a Maxwellian distribution, we can write
\begin{equation}\label{bol}
    \rho(E)=\frac{2}{\sqrt{\pi}}\frac{1}{(k_{\rm B}T_e)^{3/2}}\sqrt{E}~e^{-E/(k_{\rm B}T_e)},
\end{equation}
where $k_{\rm B}$ is the Boltzmann constant.

Equation (\ref{bol}) assumes that the free electrons are in thermal equilibrium at a temperature $T_e$, which is a common assumption in collisional-radiative models. Deviations from a Maxwellian distribution can be considered. For instance, at high densities, it may be relevant to replace the Maxwell-Boltzmann distribution with the Fermi-Dirac distribution \cite{Benredjem2022,Benredjem2024}. When hot electrons are present, such as in ultra-high-intensity laser experiments \cite{Dervieux2015}, a bi-Maxwellian distribution may be used, with one component evaluated at $T_e$ and the other at $T_{\mathrm{hot}}$ (representing the hot electrons). Moreover, electron distributions in laser-produced plasmas tend to evolve toward a super-Gaussian shape due to inverse bremsstrahlung absorption (the Langdon effect) \cite{Langdon1980}. The parameter of the super-Gaussian can be inferred from the density and temperature \cite{Sheil2025,Le2025}. Finally, including non-LTE effects for free electrons may be possible, although it remains a very challenging task. To determine the non-LTE level populations, one must solve the collisional-radiative equations, which require knowledge of the rate coefficients. If these coefficients themselves depend on the non-LTE populations, the system becomes unsolvable—even in a self-consistent manner—not to mention the considerable number of states in the continuum, due to the partial loss of quantization. An interesting alternative might be to use effective temperatures (sometimes referred to as ionization temperatures) \cite{Busquet1993,Bauche2006}. These are defined so that the LTE spectrum at $T_Z$ closely resembles the non-LTE spectrum at $T_e$. The value of $T_Z$ is chosen to ensure that the LTE ionization of the plasma at $T_Z$ matches the average non-LTE ionization at $T_e$.

\subsection{Transition arrays}
\begin{figure}[h]
\centering
\begin{tikzpicture}[scale=1.3]
\begin{scope}[shift={(-1cm,0cm)}]

\draw[thick] (.2,.3) -- (2,.3);
\draw[thick] (3,.3) -- (4.8,.3);
\draw[thick] (6.3,.3) -- (8.1,.3);

\node[left] at (5.5,.6) {$C_I$}; 
\node[right] at (2,.6) {$i$};

\draw[thick] (.2,.6) -- (2,.6);
\draw[thick] (3,.6) -- (4.8,.6);
\draw[thick] (6.3,.6) -- (8.1,.6);

\draw[thick] (.2,.9) -- (2,.9);
\draw[thick] (3,.9) -- (4.8,.9);
\draw[thick] (6.3,.9) -- (8.1,.9);

\draw[green, dashed, line width=1pt] (2.9,1.4) rectangle (4.9,2.2);
\draw[green, dashed, line width=1pt] (2.9,0.2) rectangle (4.9,1.);

\draw[green, dashed, line width=1pt] (6.2,0.2) rectangle (8.2,2.2);
\draw[green, dashed, line width=1pt] (6.2,2.9) rectangle (8.2,5.2);

\draw[thick] (.2,1.5) -- (2,1.5);
\draw[thick] (.2,1.8) -- (2,1.8) node[right] {$f$};
\draw[thick] (.2,2.1) -- (2,2.1);

\node[left] at (5.5,1.8) {$C_F$};

\draw[thick] (.2,1.5) -- (2,1.5);
\draw[thick] (3,1.5) -- (4.8,1.5);
\draw[thick] (6.3,1.5) -- (8.1,1.5);

\draw[thick] (.2,1.8) -- (2,1.8);
\draw[thick] (3,1.8) -- (4.8,1.8);
\draw[thick] (6.3,1.8) -- (8.1,1.8);

\draw[thick] (.2,2.1) -- (2,2.1);
\draw[thick] (3,2.1) -- (4.8,2.1);
\draw[thick] (6.3,2.1) -- (8.1,2.1);

\draw[thick] (.2,3) -- (2,3);
\draw[thick] (.2,3.3) -- (2,3.3);
\draw[thick] (.2,3.6) -- (2,3.6);

\draw[thick] (.2,3) -- (2,3);
\draw[thick] (3,3) -- (4.8,3);
\draw[thick] (6.3,3) -- (8.1,3);

\draw[thick] (.2,3.3) -- (2,3.3);
\draw[thick] (3,3.3) -- (4.8,3.3);
\draw[thick] (6.3,3.3) -- (8.1,3.3);

\draw[thick] (.2,3.6) -- (2,3.6);
\draw[thick] (3,3.6) -- (4.8,3.6);
\draw[thick] (6.3,3.6) -- (8.1,3.6);

\draw[thick] (.2,4.5) -- (2,4.5);
\draw[thick] (3,4.5) -- (4.8,4.5);
\draw[thick] (6.3,4.5) -- (8.1,4.5);

\draw[thick] (.2,4.8) -- (2,4.8);
\draw[thick] (3,4.8) -- (4.8,4.8);
\draw[thick] (6.3,4.8) -- (8.1,4.8);

\draw[thick] (.2,5.1) -- (2,5.1);
\draw[thick] (3,5.1) -- (4.8,5.1);
\draw[thick] (6.3,5.1) -- (8.1,5.1);

\node at (1.1, -.2) {\bf Transition};
\node at (3.9, -.2) {\bf \textcolor{blue}{Transition Array}};
\node at (7.2, -.2) {\bf \textcolor{magenta}{Super-Transition Array}};

\draw[->,line width=2pt] (1.1,.6) -- (1.1,1.8) node[midway,right]{};
\draw[->,blue, line width=2pt] (3.6,.6) -- (3.6,1.8) node[midway,right]{};

\draw[->,blue, line width=2pt] (4.2,.9) -- (4.2,2.1) node[midway,right]{};

\draw[->,magenta,line width=2pt] (6.9,.9) -- (6.9,3.6) node[midway,right]{};
\draw[->,magenta,line width=2pt] (7.5,1.5) -- (7.5,4.8) node[midway,right]{};

\node[right] at (8.3,1.2) {$\Xi_I$};
\node[right] at (8.3,4) {$\Xi_F$};

\end{scope}
\end{tikzpicture}
\caption{Hierarchy of level grouping in atomic structure. $i$ ($f$): initial (final) relativistic energy level; $C_I$ ($C_F$): initial (final) electronic configuration; $\Xi_I$ ($\Xi_F$): initial (final) superconfiguration. This diagram is relevant to excitation.}\label{Hierarchy}
\end{figure}
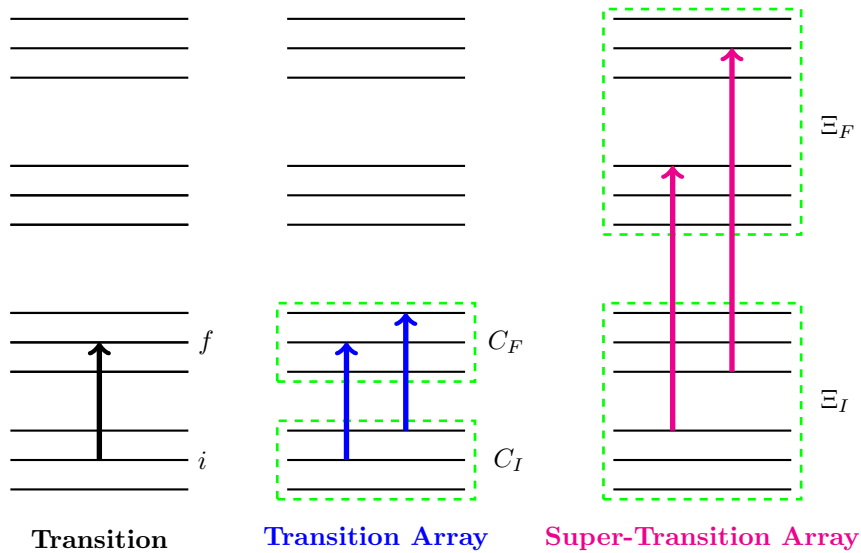

 The diagram in Figure \ref{Hierarchy} represents the hierarchy in atomic structure, going from level to configurations and superconfigurations, and from transition to transition arrays and super-transition arrays.
 
When studying excitation, we focus on transition arrays $C_I-C_F$ of the type $n_1\ell_1^{N_1}n_2\ell_2^{N_2}-n_1\ell_1^{N_1-1}n_2\ell_2^{N_2+1}$, where only the active subshells are mentioned. When we investigate ionization we deal with transition arrays of types $n_1\ell_1^{N_1}n_1\ell_2^{N_2}-n_1\ell_1^{N_1}n_1\ell_2^{N_2-1}$ and $n_1\ell_1^{N_1}n_1\ell_2^{N_2}-n_1\ell_1^{N_1-1}n_1\ell_2^{N_2}$. When studying transition arrays, we generalize the cross section and rate definitions as:
\begin{equation}\label{eq:Sigma_CC'}
   \sigma_{C_I-C_F}(E)=\sum_{i\in C_I,f\in C_F}p_i\,\sigma_{if}(E),
\end{equation}
where $p_i=g_i\exp(-E_i/k_{\rm B}T_e)/\mathcal{Z}$ is the population fraction of level $i\in C_I$, with $E_i$ and $g_i$ being the level energy and degeneracy, respectively, while $\mathcal{Z}$ is the partition function, $\mathcal{Z}=\displaystyle\sum_{j\in C_I} g_j\exp(-E_j/k_BT_e)$. Clearly, $\displaystyle \sum_{j\in C_I}p_j=1$. It is important to emphasize that, in this context, the rate is obtained by weighting the level-to-level rates with the population fractions of the initial levels and by summing over the final levels $f$.

The rate coefficient can then be written as
\begin{equation}\label{eq:q_CC'}
    q_{C_I-C_F}=\sum_{i\in C_I,f\in C_F}p_i\int_{E_{if}}^{\infty} \sigma_{if}(E)\,v\,\rho(E)\,dE
\end{equation}
or
\begin{equation}\label{eq:q_CC'2}
    q_{C_I-C_F}=\sum_{i\in C_I,f\in C_F}p_i\,q_{if}.
\end{equation}
The cross sections $\sigma_{if}$ are calculated with the Flexible Atomic Code (FAC) \cite{Gu2008}, using the distorted wave (DW) method. Here, we only consider the direct process. Actually, more than one electron of the target may be excited, leaving the atom in an unstable state. In most cases, the ion will return to its ground level by a radiative transition. But in some cases, the excited state is unstable, and a radiationless Auger transition can occur. The vacancy that is left behind by the excited electron is being filled by another electron from one of the outermost shells, while the excited electron is able to leave the ion. This process is known as excitation-autoionization (see for instance \cite{Falk1981,Mandelbaum2005}) and is not taken into account in the present study.

For each transition $i-f$, characterized by the energy $E_{if}$, we associate a specific grid representing the incident-electron energies $E$. Since the energy grids differ from one transition to another, a common energy grid encompassing all individual grids must be defined in order to express $\sigma_{C_I-C_F}$. Consequently, the cross sections $\sigma_{if}$ are interpolated onto this common energy grid. Equations \ref{eq:q_CC'} and \ref{eq:q_CC'2} should give the same rates but, due to the interpolation of the cross sections mentioned above, small differences may occur.

In the following, we investigate a set of O-like ions. For excitation, we concentrate on the two transition arrays: 
$$2s^2\,2p^4\ - 2s^1\,2p^5 \hspace{1cm} \text{and}\hspace{1cm} 2p^4\,3s^0\ - 2p^3\,3s^1$$
and concerning ionization, we consider the transition arrays:
$$2s^2\,2p^4\ - 2s^2\,2p^3\hspace{1cm} \text{and}\hspace{1cm}2s^2\,2p^4\ - 2s^1\,2p^4.$$

\begin{figure}[ht!]
    \centering
    \includegraphics[scale=.5]{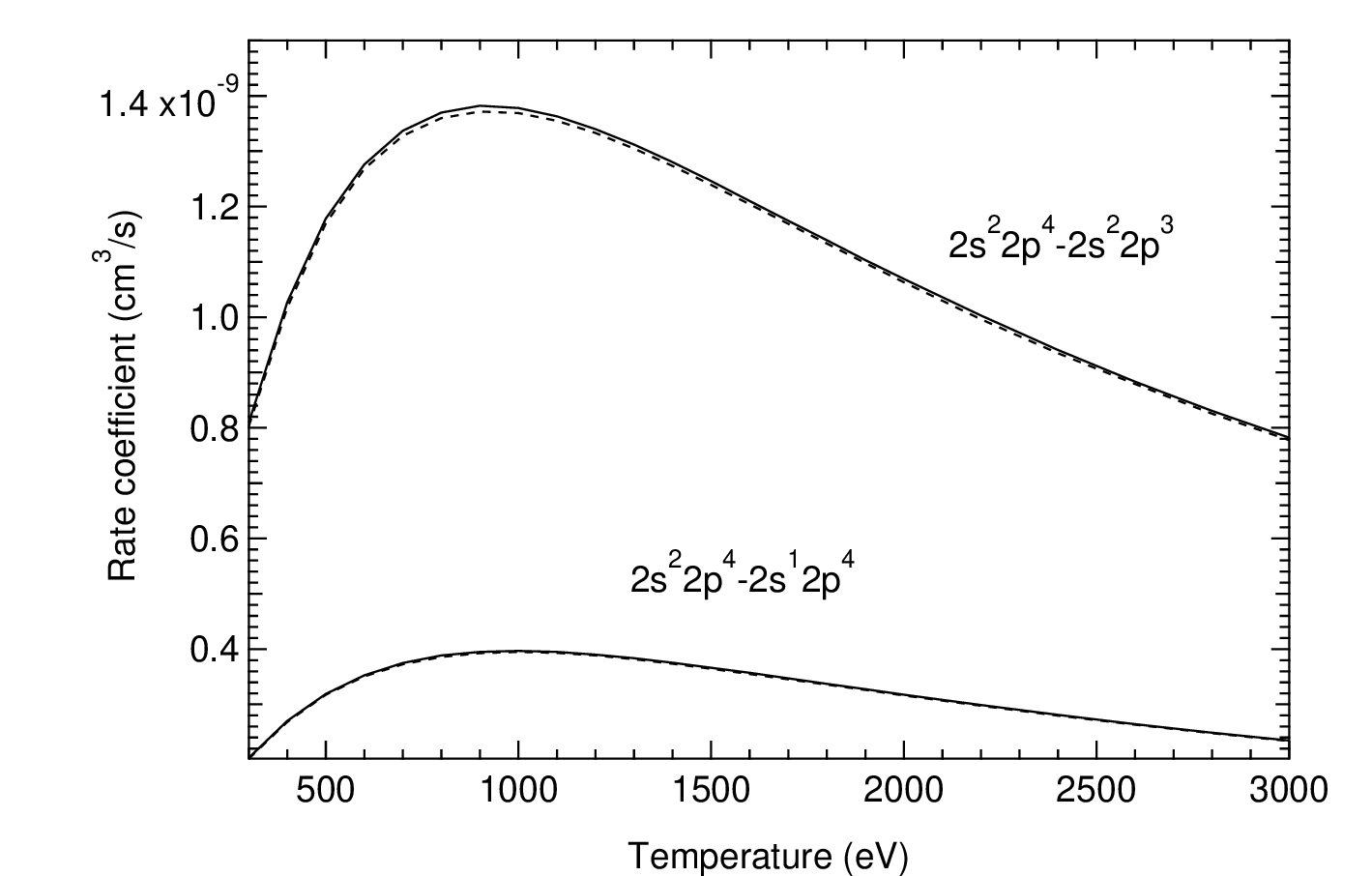}
    \caption{EII rate coefficient in the $2s^22p^4-2s^22p^3$ and $2s^22p^4-2s^12p^4$ transition arrays in O-like silicon. Solid lines: Eq. (\ref{eq:q_CC'2}), dashed lines: Eq. (\ref{eq:q_CC'}).}\label{EII-rate_SiVII}
\end{figure}

To check the accuracy of the interpolation procedure, we compare the two ionization rates (Eqs. (\ref{eq:q_CC'}) and (\ref{eq:q_CC'2})), for all O-like ions investigated. As can be seen from Figure \ref{EII-rate_SiVII}, the two calculations agree very well in the whole temperature range, showing that the interpolation of the cross sections $\sigma_{if}$ in a common energy grid is satisfactory.

\subsection{Super-transition arrays}

We also consider collisions in an STA \cite{Barshalom1989}. An STA represents an ensemble of transitions between two superconfigurations, i.e., between two ensembles of configurations (usually close in energy). Statistical methods are often used when dealing with STAs in hot plasmas, especially in non-LTE conditions, in order to make the collisional-radiative model tractable. For excitation, we concentrate on the STA $\Xi_I-\Xi_F$, where the superconfigurations $\Xi$ and $\Xi'$ are defined by:
\begin{equation}\label{eq:SC}
    \Xi_I\equiv (1)^2(2)^6=\cup\,1s^2\,2s^a\,2p^b,\ a+b=6
\end{equation}
and
\begin{equation}\label{eq:SC'}
    \Xi_F\equiv(1)^2(2)^5(3)^1=\cup\,1s^2\,2s^c\,2p^d\,3\ell^1, \ c+d=5,
\end{equation}
where $(n)$ represents the ensemble of subshells $n\ell$, $\ell$ varying from 0 to $n-1$ (Layzer complexes). Concerning ionization, we investigate the STA:
\begin{equation*}
    \Xi_I-\Xi_F\equiv (1)^2(2)^6-(1)^2(2)^5,
\end{equation*}
where $(1)^2(2)^5=\cup\,1s^2\,2s^a\,2p^b$, with $a+b=5$.

Similarly to Eq. (\ref{eq:Sigma_CC'}), we define the STA cross section: 
\begin{equation}\label{eq:Sigma_CXi-Xi'}
    \sigma_{\Xi_I-\Xi_F}(E)=\displaystyle\sum_{i\in \Xi_I,f\in \Xi_F}p_i\sigma_{if}(E),
\end{equation}
and the associate rate coefficient can be written as:

\begin{equation}\label{eq:q_xi-xip}
    q_{\Xi_I-\Xi_F}=\sum_{i\in \Xi_I,f\in \Xi_F}p_i\int_{E_{if}}^{\infty} \sigma_{if}(E)\,v\,\rho(E)\,dE
\end{equation}
or
\begin{equation}\label{eq:q_xi-xi2p}
    q_{\Xi_I-\Xi_F}=\sum_{i\in \Xi_I,f\in \Xi_F}p_i\,q_{if}.
\end{equation}
In order to facilitate comparison with the method developed in the next section, all results will be presented in Section \ref{Results} for both excitation and ionization. The method is based on two-dimensional Chebyshev polynomial expansions. As we will see, it is a powerful tool for interpolation with respect to ion charge and temperature.

\section{Chebyshev polynomial expansions}

The Chebyshev polynomials of the first kind, of order $n$ ($n\geq 0$) are defined as \cite{Abramowitz1964}:
\begin{equation*}
    T_n(x)=\cos[n\,\arccos(x)],
\end{equation*}
where $x\in [-1,1]$ is a dimensionless variable. The first two polynomials are $T_0(x)=1$ and $T_1(x)=x$. Higher-order polynomials ($n\geq 2$) can be obtained using the recurrence relation:
\begin{equation*}
    T_{n+1}(x)=2x\,T_n(x)-T_{n-1}(x).
\end{equation*}
These polynomials are orthogonal over the interval $x\in [-1,1]$, with the weight function $w(x)=1/\sqrt{1-x^2}$, i.e.,
\begin{eqnarray*}
\int_{-1}^1 \frac{T_n(x) T_p(x)}{\sqrt{1 - x^2}} \, dx =
\left\{
\begin{array}{ll}
0 & \text{if } n \neq p, \\
\pi & \text{if } n = p = 0, \\
\pi/2 & \text{if } n = p \neq 0.
\end{array}
\right.
\end{eqnarray*}

\noindent The polynomials reach their extrema, $\pm 1$, $n$ times in the interval $[-1, 1]$, and their roots are given by
\begin{equation*}
    x_k = \cos\left(\frac{k+1/2}{n} \pi \right), \quad k = 0, 1, \dots, n-1.
\end{equation*}

To approximate a function $f(x)$ defined on $[-1, 1]$, a linear combination of Chebyshev polynomials can be used:
\begin{equation}\label{eq:1D-expansion}
    f(x) = \sum_{n=0}^N c_n T_n(x),
\end{equation}
where the coefficients $c_n$ are given by
\begin{eqnarray*}
    c_0 &=& \frac{1}{\pi} \int_{-1}^1 f(x) \frac{1}{\sqrt{1-x^2}} \, dx\\
    c_n &=& \frac{2}{\pi} \int_{-1}^1 f(x) \frac{T_n(x)}{\sqrt{1-x^2}} \, dx \quad n \geq 1.
\end{eqnarray*}
The extension to two dimensions is achieved by employing the product of Chebyshev polynomials, \textit{i.e.}, one for each dimension. A smooth function $f(x,y)$, defined on $[-1, 1] \otimes [-1, 1]$ can be expanded in terms of products of two Chebyshev polynomials:
\begin{equation*}
    f(x, y) = \sum_{n=0}^N \sum_{p=0}^P c_{np} T_n(x) T_p(y),
\end{equation*}
and the Chebyshev coefficients $c_{np}$ are given by
\begin{eqnarray*}
    c_{np} = \frac{(2-\delta_{n0})(2-\delta_{p0})}{\pi^2} \int_{-1}^1 \int_{-1}^1 f(x, y) \frac{T_n(x)}{\sqrt{1-x^2}} \frac{T_p(y)}{\sqrt{1-y^2}} \, dx \, dy,
\end{eqnarray*}
where $\delta_{n0}$ is the Kronecker delta symbol.

Due to the form of the system of linear equations to be solved, these coefficients can be obtained by inverting Vandermonde matrices \cite{Gohberg1994,Eisinberg1998,Eisinberg2006}, for example, using the Fadeev-Leverrier-Souriau algorithm \cite{LeVerrier1840,Souriau1948,Fadeev1972,Hou2002}. However, fitting a polynomial to data can present challenges. The Runge phenomenon (analogous to the Gibbs phenomenon in Fourier series) refers to oscillations near the endpoints of an interval when using high-degree polynomial interpolation with equally spaced points \cite{Runge1901}. As the number of points increases, the approximation near the edges deteriorates. This issue is mitigated by choosing non-equally spaced interpolation points.

In practice, the temperature $T$ and ion charge $z$ must be normalized to the interval $[-1, 1]$ to ensure compatibility with Chebyshev polynomials. The respective normalized variables $\bar{T}$ and $\bar{z}$ are defined as follows:
\begin{equation}\label{eq:normalize-T}
    \bar{T} = \frac{2(T - T_{\text{min}})}{T_{\text{max}} - T_{\text{min}}} - 1,
\end{equation}
\begin{equation}\label{eq:normalize-z}
    \bar{z} = \frac{2(z - z_{\text{min}})}{z_{\text{max}} - z_{\text{min}}} - 1,
\end{equation}
where $T_{\rm max\,(min)}$ represents the maximum (minimum) temperature, and $z_{\rm max\,(min)}$ the maximum (minimum) ion charge considered. When $T=T_{\rm max\,(min)}$, we have $\bar{T}=1\,(-1)$. Similarly, when $z=z_{\rm max\,(min)}$, we have $\bar{z}=1\,(-1)$. To recover the original values of $T$ and $z$ from $\bar{T}$ and $\bar{z}$, the inverse transformations:
\begin{equation*}
    T = \frac{(\bar{T} + 1)(T_{\text{max}} - T_{\text{min}})}{2} + T_{\text{min}},
\end{equation*}
\begin{equation*}
    z = \frac{(\bar{z} + 1)(z_{\text{max}} - z_{\text{min}})}{2} + z_{\text{min}}
\end{equation*}
can be carried out.

For our applications, we choose:
\begin{enumerate}[label=\alph*)]
    \item $z_{\text{min}}$=6 and $z_{\text{max}}$=28, corresponding to Si$^{6+}$ and Kr$^{28+}$, respectively, and
    \item $T_{\text{min}}$=300 (50) eV, corresponding to ionization (excitation), and $T_{\text{max}}$=3000 eV for both excitation and ionization.
\end{enumerate}
We restricted ourselves to five charge states for ionization and seven for excitation. These values allow for a very satisfactory fit with Chebyshev polynomials, as can be seen in figures \ref{EII_2p4-2p3}-\ref{EII_SC1-SC2}.

The fitting procedure is applied to the logarithm of the rate data. Specifically, we define
\begin{equation}\label{eq:qbar}
    \bar{q}(\bar{T},\bar{z})=\log_{10}q(T,z),
\end{equation}
for both transition arrays and STAs. This transformation improves numerical stability and enables the model to capture variations over several orders of magnitude. The original rate values can be recovered via the simple relation:
\begin{equation*}
    q(T,z) = 10^{\log_{10}\bar{q}(\bar{T},\bar{z})}.
\end{equation*}
For discrete data, the coefficients $c_{np}$ are computed by minimizing the mean squared error (MSE) between the logarithm of the calculated and fitted rates:
\begin{equation*}
    \text{MSE} = \frac{1}{M} \sum_{m=0}^{M-1} \left[ \bar{q}_m - \sum_{n=0}^N \sum_{p=0}^P c_{np} T_n(\bar{T}_i) T_p(\bar{z}_i) \right]^2,
\end{equation*}
where $m$ indexes the $M$ data points and $n$ and $p$ denote the degrees of the Chebyshev polynomials in the $\bar{T}$ and $\bar{z}$ dimensions, ranging from 0 to $N$ and from 0 to $P$, respectively

Due to their orthogonality, Chebyshev polynomials offer efficient and accurate approximations, particularly minimizing edge errors in the fitting domain. They are well-suited for smooth functions without rapid oscillations. In contrast, Hermite polynomials are more appropriate for functions with high-frequency features. Both methods provide flexible and effective tools for representing complex distributions \cite{Turan2024}.

\section{Results}\label{Results}

\subsection{Ionization in transitions arrays}

The ionization rate is defined by Eq. (\ref{eq:q_CC'2}), where each $q_{if}$ corresponds to a transition between two relativistic energy levels ($jj$ coupling) and is derived from the cross section $\sigma_{if}$ obtained using the FAC code. Here, $C_I \equiv 1s^2 2s^2 2p^4$ (denoted $2p^4$ in the following) and $C_F \equiv 1s^2 2s^2 2p^3$ or $1s^2 2s^1 2p^4$ (denoted respectively $2p^3$ and $2s^1 2p^4$). Chebyshev coefficients $c_{np}$ are calculated by fitting the data for the O-like ions Si$^{6+}$, Ar$^{10+}$, Ti$^{14+}$, Ge$^{24+}$ and Kr$^{6+}$, across a temperature range of 300$-$3000 eV.

\begin{figure}[ht!]
    \centering
    \includegraphics[scale=.5]{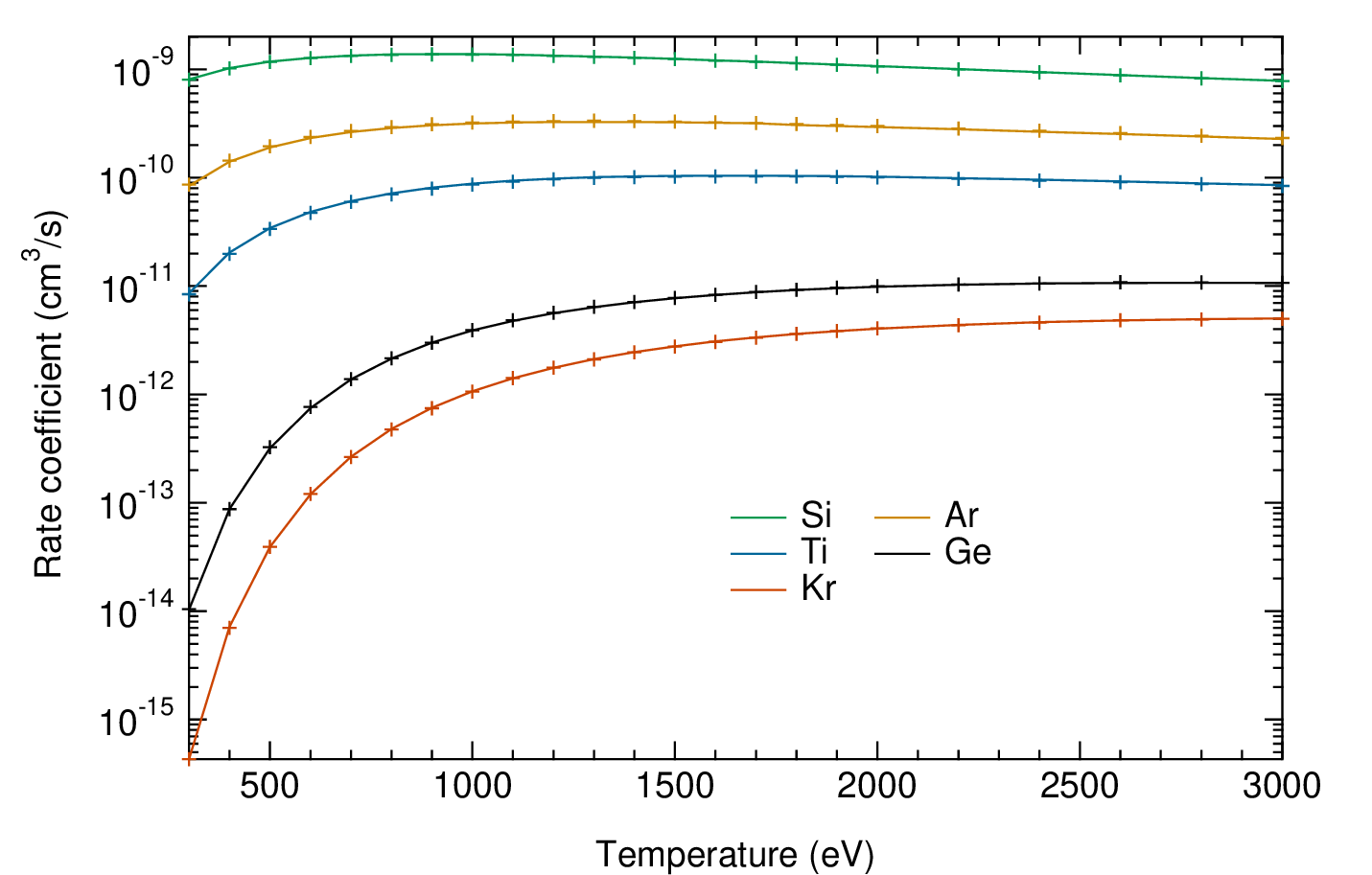}
    \caption{Ionization rate coefficient for the $2p^4-2p^3$ transition array in O-like ions. Solid lines: direct calculations using Eq. (\ref{eq:q_CC'}), cross markers: two-dimension Chebyshev polynomial fit.}\label{EII_2p4-2p3}
\end{figure}

Figure \ref{EII_2p4-2p3} shows the EII rate for the $2p^4-2p^3$ transition array. The rate is plotted as a function of temperature for five O-like ions with net positive charges of 6, 10, 14, 24, and 28. The rate decreases with increasing ion charge, while the temperature dependence remains qualitatively similar for all ions. The Chebyshev polynomial fit accurately reproduces very well the computed rates using polynomials $T_n(\bar{T})$ ($0\leq n\leq 17$) and $T_p(\bar{z})$ ($0\leq p\leq 3$), yielding $4\times 18$ coefficients $c_{np}$ (see Table \ref{Cheb-coef-EII-2$p^4-2p^3$}). The procedure used to compute the rate for any temperature in [300,3000] eV and any O-like ion with net charge in [6,28] consists in: 
\begin{enumerate}[label=(\alph*)]
\item Normalizing $T$ and $z$ using Eqs. (\ref{eq:normalize-T}) and (\ref{eq:normalize-z}), yielding $\bar{T}$ and $\bar{z}$,
\item Computing $\bar{q}$, using Eq. (\ref{eq:qbar}): \[\bar{q}_{C_I-C_F}(\bar{T},\bar{z})=\sum_{n=0}^{N} \sum_{p=0}^{P}c_{n,p}T_n(\bar{T})T_p(\bar{z}).\]
This is achieved through a fitting procedure that determines the Chebyshev coefficients $c_{np}$, ensuring a high degree of accuracy with respect to the mean squared error (MSE).
\item Recovering the rate coefficient (in cm$^3$/s) as $\displaystyle q_{C_I-C_F}=10^{\bar{q}_{C_I-C_F}}$.
\end{enumerate}

\begin{table}[ht!]\caption{Chebyshev coefficients, $c_{np}$, for ionization in the 2$p^4-2p^3$ transition array.}\label{Cheb-coef-EII-2$p^4-2p^3$}
\begin{center}
\begin{tabular}{ c| c | c | c | c}
\hline\hline
$n\;\downarrow\;p \rightarrow$ & 0 & 1 & 2 & 3\\
\hline
0 & -1.06129$\times 10^1$ & -1.54235 & 7.87700$\times 10^{-2}$ & -4.20639$\times 10^{-2}$\\
1 & 5.58029$\times 10^{-1}$ & 7.98115$\times 10^{-1}$ & 2.06388$\times 10^{-1}$ & -4.02129$\times 10^{-2}$\\
2 & -4.44599$\times 10^{-1}$ & -3.42177$\times 10^{-1}$ & 5.00029$\times 10^{-2}$ & -3.68038$\times 10^{-2}$\\
3 & 2.03552$\times 10^{-1}$ & 2.34639$\times 10^{-1}$ & 1.04379$\times 10^{-1}$ & -2.88719$\times 10^{-2}$\\
4 & -1.25024$\times10^{-1}$ & -7.99232$\times 10^{-2}$ & 3.77901$\times 10^{-2}$ & -1.87027$\times 10^{-2}$\\
5 & 5.08201$\times 10^{-2}$ & 6.35216$\times 10^{-2}$ & 2.94419$\times 10^{-2}$ & -8.77172$\times 10^{-3}$\\
6 & -2.74920$\times 10^{-2}$ & -2.98393$\times 10^{-2}$ & -1.13269$\times 10^{-2}$ & 1.98965$\times 10^{-3}$\\
7 & 2.19953$\times 10^{-2}$ & 2.43408$\times 10^{-4}$ & -3.15560$\times 10^{-2}$ & 1.12374$\times 10^{-2}$\\
8 & 4.63419$\times 10^{-3}$ & -3.00965$\times 10^{-2}$ & -5.46717$\times 10^{-2}$ & 1.85975$\times 10^{-2}$\\
9 & 1.87540$\times 10^{-2}$ & -2.41377$\times 10^{-2}$ & -6.48405$\times 10^{-2}$ & 2.27470$\times 10^{-2}$\\
10 & 1.36188$\times 10^{-2}$ & -3.09526$\times 10^{-2}$ & -6.81196$\times 10^{-2}$ & 2.36961$\times 10^{-2}$\\
11 & 1.60540$\times 10^{-2}$ & -2.70364$\times 10^{-2}$ & -6.28662$\times 10^{-2}$ & 2.29417$\times 10^{-2}$\\
12 & 1.20889$\times 10^{-2}$ & -2.34593$\times 10^{-2}$ & -5.10942$\times 10^{-2}$ & 1.89064$\times 10^{-2}$\\
13 & 1.02643$\times 10^{-2}$ & -1.79633$\times 10^{-2}$ & -3.77954$\times 10^{-2}$ & 1.50728$\times 10^{-2}$\\
14 & 6.39619$\times 10^{-3}$ & -1.15549$\times 10^{-2}$ & -2.31351$\times 10^{-2}$ & 9.54846$\times 10^{-3}$\\
15 & 4.58860$\times 10^{-3}$ & -7.71375$\times 10^{-3}$ & -1.30652$\times 10^{-2}$ & 6.56337$\times 10^{-3}$\\
16 & 1.97863$\times 10^{-3}$&-3.33047$\times 10^{-3}$&-5.00320$\times 10^{-3}$&2.84116$\times 10^{-3}$\\
17 & 1.00572$\times 10^{-3}$&-1.55603$\times 10^{-3}$&-1.79323$\times 10^{-3}$&1.37074$\times 10^{-3}$\\
\hline\hline
\end{tabular}
\end{center}
\end{table}

\begin{figure}[ht!]
    \centering
    \includegraphics[scale=.5]{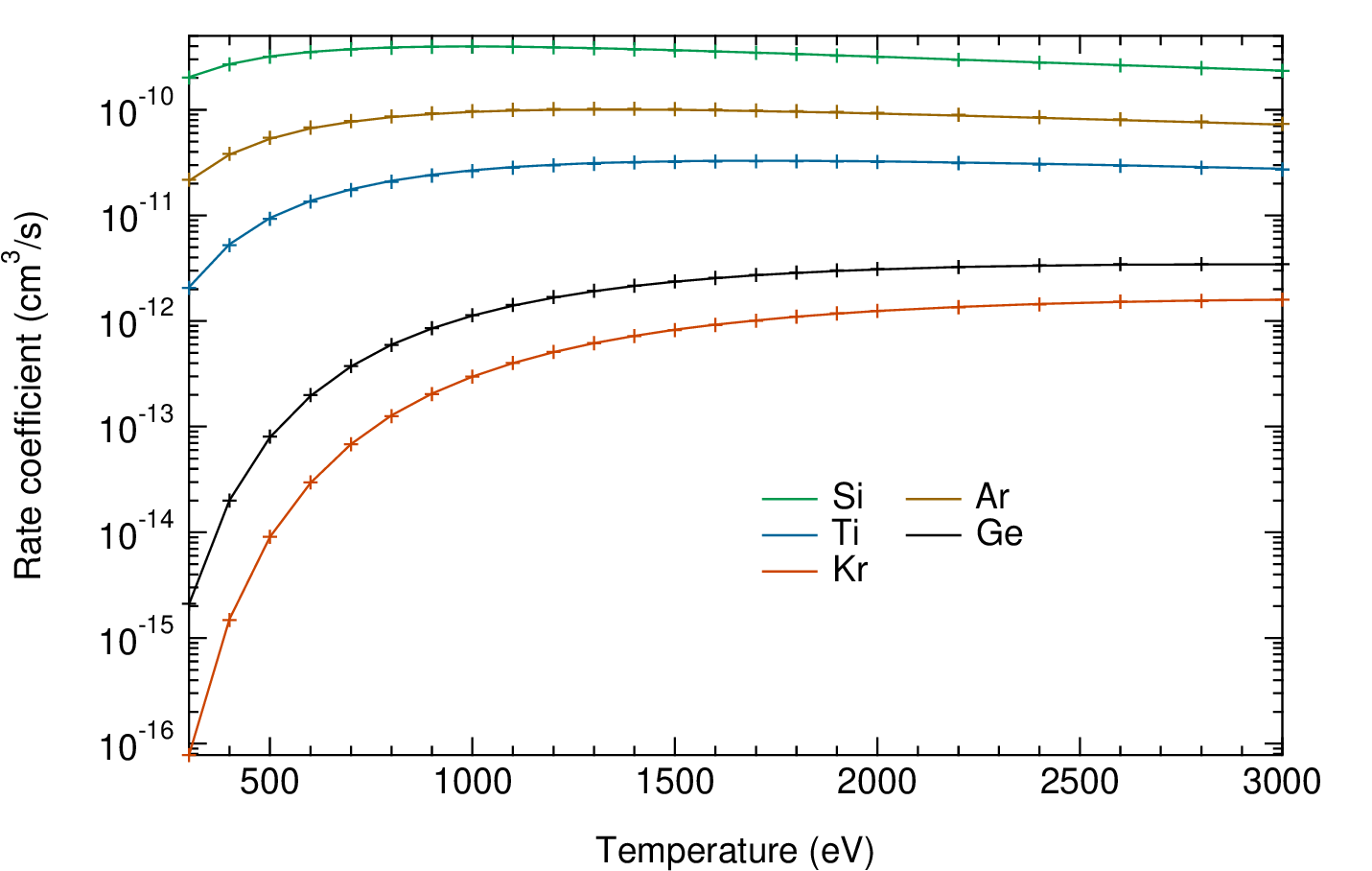}
    \caption{Ionization rate coefficient for the $2p^4-2s^12p^4$ transition array in O-like ions. Solid lines: direct calculations using Eq. (\ref{eq:q_CC'}), cross markers: two-dimension Chebyshev polynomial fit.}\label{EII_2p4-2s12p4}
\end{figure}

Figure \ref{EII_2p4-2s12p4} represents the EII rate for the $2p^4-2s^12p^4$ transition array. The behavior is similar to that of the $2p^4-2p^3$ array (see Fig. \ref{EII_2p4-2p3}). The obtained Chebyshev coefficients $c_{np}$ are provided in Table \ref{Cheb-coef-EII-$2p^4-2s^12p^4$}. The same procedure allows calculation of the rate for any temperature and ion charge within the specified ranges. The rate values for $2p^4-2s^12p^4$ are lower than those for $2p^4-2p^3$, with a relative magnitude between the two transition arrays ranging from 3 to 4 over all temperatures and ions. The difference is partially due to the energy difference between the two upper configurations.\\
\indent Now let us focus on excitation by electron impacts between two configurations.

\begin{table}[ht!]\caption{Chebyshev coefficients, $c_{np}$, for ionization in the $2p^4-2s^12p^4$ transition array.}\label{Cheb-coef-EII-$2p^4-2s^12p^4$}
\begin{center}
\begin{tabular}{c|c|c|c|c}
\hline\hline
$n\;\downarrow\;p \rightarrow$ & 0 & 1 & 2 & 3\\
\hline
0 & -1.11616$\times10^1$ & -1.53504 & 5.39080$\times10^{-2}$ & -4.34142$\times10^{-2}$\\
1 & 5.89237$\times10^{-1}$ & 8.62841$\times10^{-1}$ & 1.88432$\times10^{-1}$ & -4.69724$\times10^{-2}$\\
2 & -4.95571$\times10^{-1}$ & -3.27189$\times10^{-1}$ & 2.97825$\times10^{-2}$ & -4.34706$\times10^{-2}$\\
3 & 2.00920$\times10^{-1}$ & 2.66676$\times10^{-1}$ & 9.05926$\times10^{-2}$ & -3.36992$\times10^{-2}$\\
4 & -1.44052$\times10^{-1}$ & -6.80846$\times10^{-2}$ & 2.76780$\times10^{-2}$ & -2.22649$\times10^{-2}$\\
5 & 5.01439$\times10^{-2}$ & 7.18794$\times10^{-2}$ & 2.54706$\times10^{-2}$ & -9.70456$\times 10^{-3}$\\
6 & -2.70811$\times10^{-2}$ & -3.42886$\times10^{-2}$ & -1.02152$\times10^{-2}$ & 3.22757$\times 10^{-3}$\\
7 & 3.13095$\times10^{-2}$ & -1.07383$\times10^{-2}$ & -2.53862$\times10^{-2}$ & 1.42973$\times10^{-2}$\\
8 & 1.69590$\times10^{-2}$ & -4.82275$\times10^{-2}$ & -4.48935$\times10^{-2}$ & 2.29836$\times10^{-2}$\\
9 & 3.45746$\times10^{-2}$ & -4.56135$\times10^{-2}$ & -5.28259$\times10^{-2}$ & 2.80873$\times10^{-2}$\\
10 & 2.95731$\times10^{-2}$ & -5.33378$\times10^{-2}$ & -5.56207$\times10^{-2}$ & 2.91312$\times10^{-2}$\\
11 & 3.03318$\times10^{-2}$ & -4.66754$\times10^{-2}$ & -5.12082$\times10^{-2}$ & 2.72742$\times10^{-2}$\\
12 & 2.32603$\times10^{-2}$ & -3.88913$\times10^{-2}$ & -4.13994$\times10^{-2}$ & 2.19523$\times10^{-2}$\\
13 & 1.77428$\times10^{-2}$ & -2.82906$\times10^{-2}$ & -3.06326$\times10^{-2}$ & 1.66244$\times10^{-2}$\\
14 & 1.05655$\times10^{-2}$ & -1.72837$\times10^{-2}$ & -1.84425$\times10^{-2}$ & 1.00052$\times10^{-2}$\\
15 & 6.04300$\times 10^{-3}$ & -9.67233$\times 10^{-3}$ & -1.04201$\times10^{-2}$ & 5.96682$\times10^{-3}$\\
16 & 2.22114$\times10^{-3}$ & -3.57739$\times10^{-3}$ & -3.76995$\times10^{-3}$ & 2.18845$\times10^{-3}$\\
17 & 8.38359$\times10^{-4}$ & -1.31825$\times10^{-3}$ & -1.40796$\times10^{-3}$ & 9.40161$\times10^{-4}$\\
\hline\hline
\end{tabular}
\end{center}
\end{table}

\subsection{Excitation in transitions arrays}
We calculate the rate coefficient for the two distinct transition arrays: $2s^22p^4-2s^22p^33s^1$ ($2p-3s$ transition type) and $2s^22p^4-2s^12p^5$ ($2s-2p$ transition type). The lower bound of the temperature range is now 50 eV, increasing in steps of 50 eV up to 300 eV. From 300 eV to 2000 eV, the step size is 100 eV, and for temperatures above 2000 eV, the step size increases to 200 eV. Here, we investigate seven O-like ions: Si$^{6+}$, Ar$^{10+}$, Ti$^{14+}$, Fe$^{18+}$, Cu$^{21+}$, Ge$^{24+}$ and Kr$^{28+}$. As for ionization, we first calculate the cross section using the FAC code. Then, assuming a Maxwellian distribution for electron projectiles, we calculate the rate.

\begin{figure}[ht!]
    \centering
    \includegraphics[scale=.5]{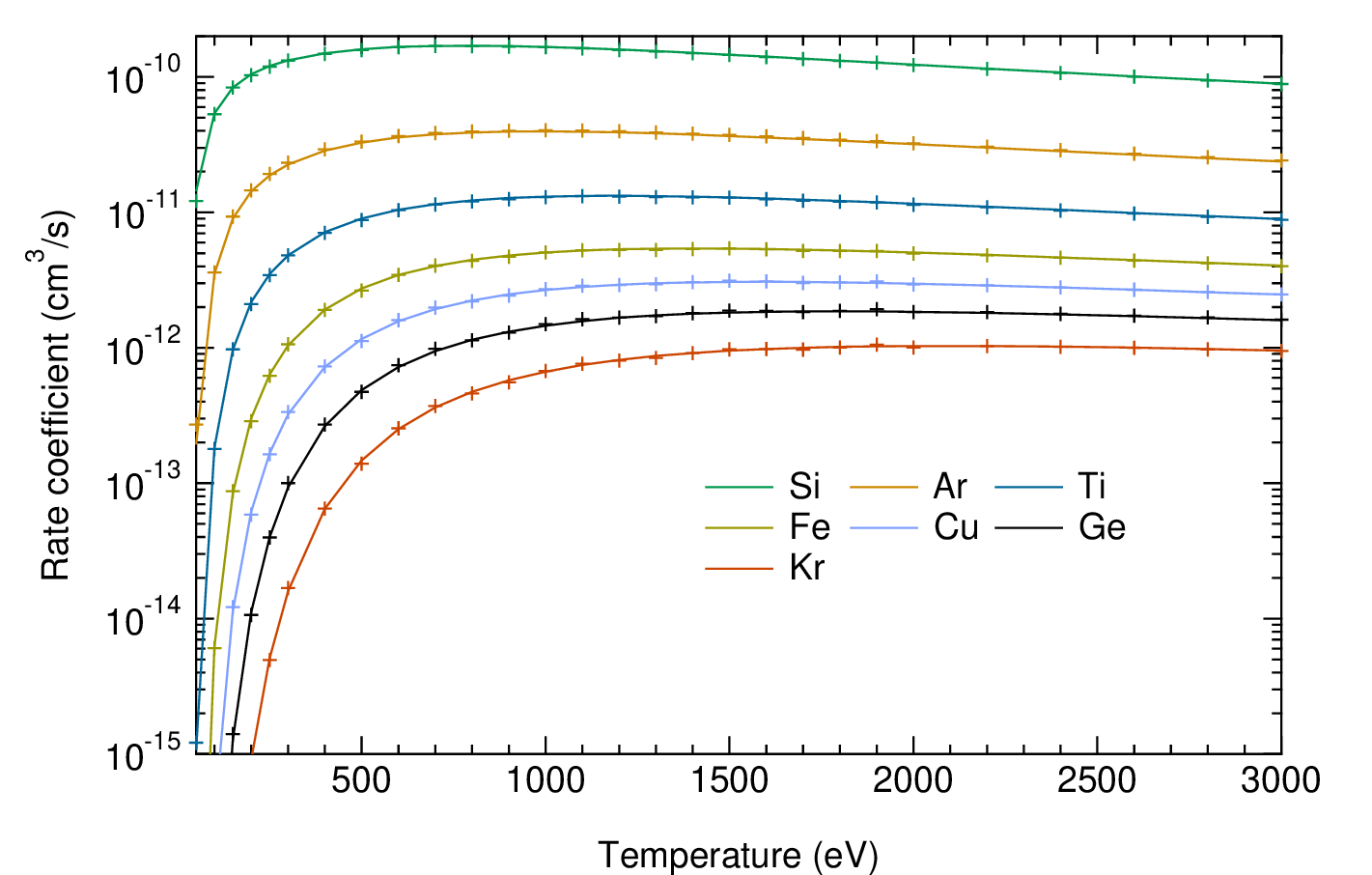}
    \caption{Excitation rate coefficient in the $2p^4-2p^33s^1$ transition array in O-like ions. Solid lines: direct calculations using Eq. (\ref{eq:q_CC'}), cross markers: fit using Chebyshev polynomials.}\label{EIE_2p-3s_extended}
\end{figure}

Figure \ref{EIE_2p-3s_extended} shows the EIE rate for the $2p-3s$ transition array. Similarly to ionization, we also present the fit of the calculated rates using two-dimension expansions in terms of products of two Chebyshev polynomials. To achieve a good agreement between the direct calculation and the corresponding fit, we use $4\times 18$ products of Chebyshev polynomials of the form $T_n(\bar{T})T_p(\bar{z})$. The agreement between the two calculations is very satisfactory. The Chebyshev coefficients $c_{np}$ are given in Table \ref{Cheb-coef-EIE-2p-3s}.

\begin{table}[ht!]\caption{Chebyshev coefficients, $c_{np}$, for excitation in the $2p-3s$ transition array.}\label{Cheb-coef-EIE-2p-3s}
\begin{center}
\begin{tabular}{c|c|c|c|c}
\hline\hline
$n\;\downarrow\;p \rightarrow$ & 0 & 1 & 2 & 3\\
\hline
0 & -1.36371$\times 10^1$ & -3.64576 & -2.00684$\times10^{-1}$ & 2.06928$\times10^{-1}$\\
1 & -2.39099 & -2.30492 & -2.27801$\times10^{-1}$ & 4.33370$\times10^{-1}$\\	
2 & -4.34628 & -4.37372 & -5.51413$\times10^{-1}$ & 4.05890$\times10^{-1}$\\
3 & -1.71551 & -1.79363 & -1.85319$\times10^{-1}$ & 3.04897$\times10^{-1}$\\	
4 & -2.47712 & -2.51111 & -3.14900$\times10^{-1}$ & 2.33458$\times10^{-1}$\\	
5 & -3.71518$\times10^{-1}$ & -3.89237$\times10^{-1}$ & -3.32000$\times10^{-2}$ & 9.74695$\times10^{-2}$\\	
6 & -3.86543$\times10^{-1}$ & -3.88542$\times10^{-1}$ & -5.44277$\times10^{-2}$ & 8.59621$\times10^{-3}$\\	
7 & 1.19300 & 1.21349 & 1.48936$\times10^{-1}$ & -1.17645$\times10^{-1}$\\	
8 & 1.31272 & 1.34469 & 1.54782$\times10^{-1}$ & -1.81920$\times10^{-1}$\\	
9 & 2.21468 & 2.25896 & 2.69140$\times10^{-1}$ & -2.56384$\times10^{-1}$\\	
10 & 2.06277 & 2.10900 & 2.47068$\times10^{-1}$ & -2.63545$\times10^{-1}$\\	
11 & 2.25665 & 2.30356 & 2.72963$\times10^{-1}$ & -2.69213$\times10^{-1}$\\	
12 & 1.76440 & 1.80295 & 2.11958$\times10^{-1}$ & -2.21099$\times10^{-1}$\\	
13 & 1.49611 & 1.52813 & 1.80797$\times10^{-1}$ & -1.80055$\times10^{-1}$\\	
14 & 9.16787$\times10^{-1}$ & 9.36461$\times10^{-1}$ & 1.10210$\times10^{-1}$ & -1.14083$\times10^{-1}$\\	
15 & 5.95014$\times10^{-1}$ & 6.08321$\times10^{-1}$ & 7.19648$\times10^{-2}$ & -7.16270$\times10^{-2}$\\	
16 & 2.31825$\times10^{-1}$ & 2.36737$\times10^{-1}$ & 2.78623$\times10^{-2}$ & -2.87782$\times10^{-2}$\\
17 & 1.03203$\times10^{-1}$ & 1.05741$\times10^{-1}$ & 1.25117$\times10^{-2}$ & -1.22859$\times10^{-2}$\\
\hline\hline
\end{tabular}
\end{center}
\end{table}

\begin{figure}[ht!]
    \centering
    \includegraphics[scale=.5]{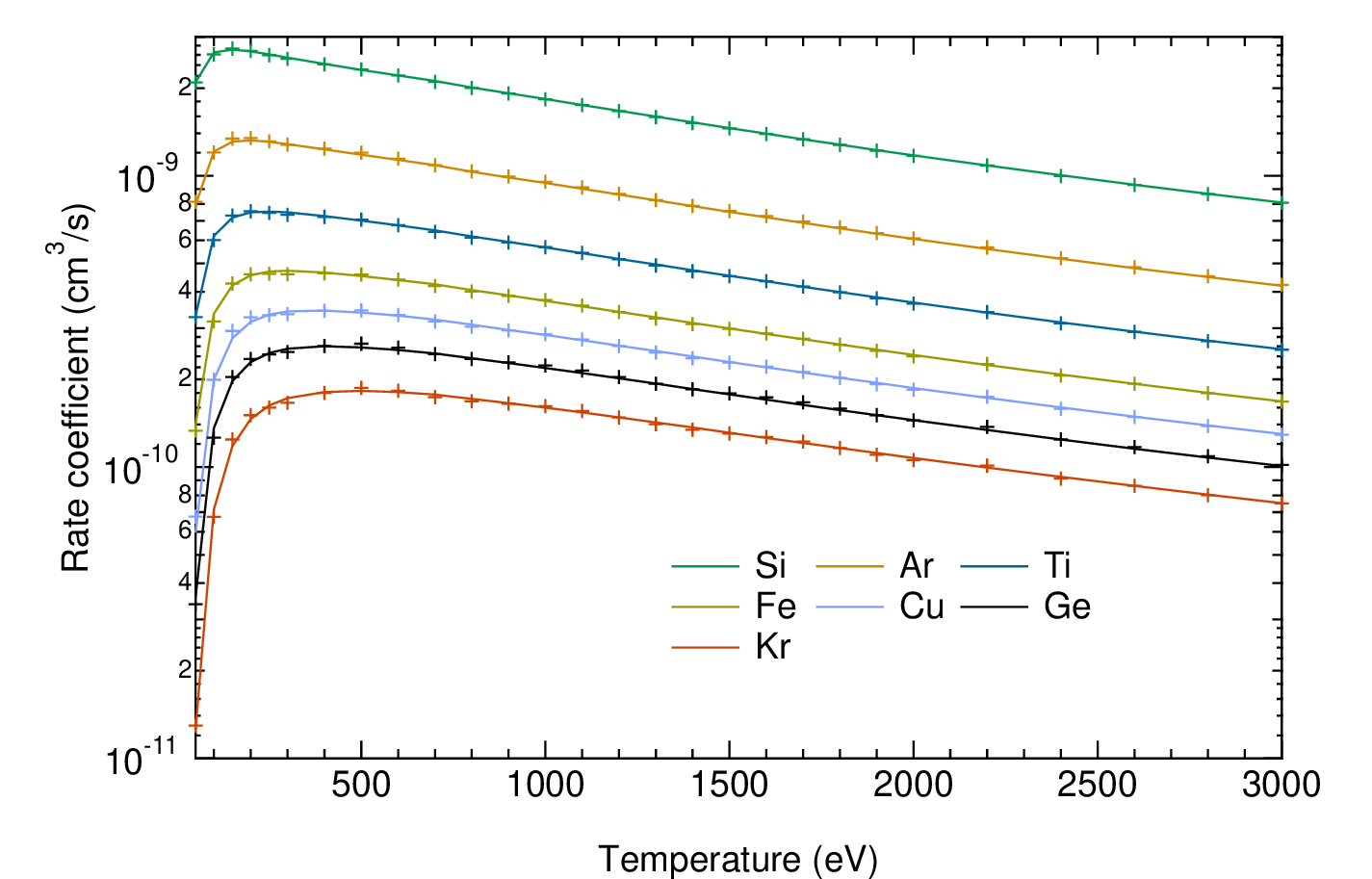}
    \caption{Excitation rate coefficient in the transition array $2p^4-2s^12p^5$ in O-like ions. Solid lines: direct calculations using Eq. (\ref{eq:q_CC'}), cross markers: fit using Chebyshev polynomials.}\label{EIE_2s-2p_extended}
\end{figure}

Figure \ref{EIE_2s-2p_extended} illustrates the variation of the EIE rate for the $2s-2p$ transition array. The procedure follows the same approach as the $2p-3s$ transition array, with identical temperature and ion charge ranges. The Chebyshev coefficients $c_{np}$ are provided in Table \ref{Cheb-coef-EIE-2s-2p}. A comparison with the previous case (Fig. \ref{EIE_2p-3s_extended}) reveals that the ratio of the maximum rates of the two transition arrays for each ion ranges between 16 and 170.\\

\begin{table}[ht!]\caption{Chebyshev coefficients, $c_{np}$, for excitation in the $2s-2p$ transition array.}\label{Cheb-coef-EIE-2s-2p}
\begin{center}
\begin{tabular}{c|c|c|c|c}
\hline\hline
$n\;\downarrow\;p \rightarrow$ & 0 & 1 & 2 & 3\\
\hline
0 & -9.68545&-7.13518$\times10^{-1}$&3.96736$\times10^{-2}$&-1.06229$\times10^{-2}$\\
1 & -5.81465$\times10^{-1}$ & -1.75676$\times10^{-1}$ & -1.45415$\times10^{-2}$ & 2.88991$\times10^{-3}$\\
2 & -4.53569$\times10^{-1}$ & -3.46993$\times10^{-1}$ & -3.64651$\times10^{-2}$ & 4.31319$\times10^{-3}$\\
3 & -2.18255$\times10^{-1}$ & -1.32898$\times10^{-1}$ & -1.25553$\times10^{-2}$ & 1.90927$\times10^{-3}$\\
4 & -2.73289$\times10^{-1}$ & -2.02012$\times10^{-1}$ & -2.03633$\times10^{-2}$ & 2.55014$\times10^{-3}$\\
5 & -5.03660$\times10^{-2}$ & -2.49514$\times10^{-2}$ & -2.63609$\times10^{-3}$ & 4.20455$\times10^{-4}$\\
6 & -4.09637$\times10^{-2}$ & -3.45947$\times10^{-2}$ & -3.22050$\times10^{-3}$ & 3.09609$\times10^{-4}$\\
7 & 1.36039$\times10^{-1}$ & 9.78758$\times10^{-2}$ & 9.52477$\times10^{-3}$ & -1.22024$\times10^{-3}$\\
8 & 1.51655$\times10^{-1}$ & 1.02223$\times10^{-1}$ & 1.04371$\times10^{-2}$ & -1.49604$\times10^{-3}$\\
9 & 2.55759$\times10^{-1}$ & 1.77837$\times10^{-1}$ & 1.75854$\times10^{-2}$ & -2.28073$\times10^{-3}$\\
10 & 2.37413$\times10^{-1}$&1.62973$\times10^{-1}$ & 1.64190$\times10^{-2}$ & -2.27692$\times10^{-3}$\\
11 & 2.60829$\times10^{-1}$ & 1.80391$\times10^{-1}$ & 1.79440$\times10^{-2}$ & -2.31478$\times10^{-3}$\\
12 & 2.03422$\times10^{-1}$ & 1.39819$\times10^{-1}$ & 1.40608$\times10^{-2}$ & -1.91877$\times10^{-3}$\\
13 & 1.72839$\times10^{-1}$ & 1.19518$\times10^{-1}$ & 1.19269$\times10^{-2}$ & -1.49545$\times10^{-3}$\\
14 & 1.05997$\times10^{-1}$ & 7.26670$\times10^{-2}$ & 7.33100$\times10^{-3}$ & -9.43641$\times10^{-4}$\\
15 & 6.90087$\times10^{-2}$ & 4.76264$\times10^{-2}$ & 4.76467$\times10^{-3}$ & -5.47181$\times10^{-4}$	\\
16 & 2.68005$\times10^{-2}$ & 1.83440$\times10^{-2}$ & 1.88251$\times10^{-3}$ & -2.13653$\times10^{-4}$	\\
17 & 1.22453$\times10^{-2}$ & 8.32156$\times10^{-3}$ & 8.24676$\times10^{-4}$ & -5.15302$\times10^{-5}$\\
\hline\hline
\end{tabular}
\end{center}
\end{table}

In the following, we investigate the rate variation in STAs. The number of transitions between relativistic energy levels is much larger than in transition arrays.

\subsection{Excitation and ionization in super-transitions arrays}
\begin{table}[ht!]\caption{Chebyshev coefficients, $c_{np}$, for excitation in the STA $(1)^2(2)^6-(1)^2(2)^5(3)^1$.}\label{Cheb-coef-EIE-SC1-SC2}
\begin{center}
\begin{tabular}{c|c|c|c|c}
\hline\hline
$n\;\downarrow\;p \rightarrow$ & 0 & 1 & 2 & 3\\
\hline
0 & -1.26477$\times 10^{1}$ &-3.86272 & -3.42948$\times 10^{-1}$ & -2.15910$\times 10^{-2}$\\
1 & -3.50846 & -2.99456 & -3.56586$\times 10^{-1}$ & -1.04175$\times 10^{-2}$\\
2 & -5.43872 & -5.15683 & -6.68698$\times 10^{-1}$ & -8.30189$\times 10^{-3}$\\
3 & -2.50420 & -2.31917&-2.84305$\times 10^{-1}$ & -8.38387$\times 10^{-3}$\\
4 & -3.09815 & -2.93533 & -3.77776$\times 10^{-1}$ & -4.30751$\times 10^{-3}$\\
5 & -6.27629$\times 10^{-1}$ & -5.69525$\times 10^{-1}$ & -6.86824$\times 10^{-2}$ & -3.12485$\times 10^{-3}$\\
6 & -3.99108$\times 10^{-1}$ & -3.88037$\times 10^{-1}$ & -5.13022$\times 10^{-2}$ & 4.77957$\times 10^{-4}$\\
7 & 1.49501&1.41212 & 1.80226$\times 10^{-1}$ & 2.25380$\times 10^{-3}$\\
8 & 1.81016 & 1.69411 & 2.13821$\times 10^{-1}$ & 4.68344$\times 10^{-3}$\\
9 & 2.88044 & 2.70762 & 3.43603$\times 10^{-1}$ & 5.69882$\times 10^{-3}$\\
10 & 2.77777 & 2.60582 & 3.29837$\times 10^{-1}$ & 6.47641$\times 10^{-3}$\\
11 & 2.95392 & 2.77470 & 3.51806$\times 10^{-1}$ & 6.06635$\times 10^{-3}$\\
12 & 2.36071 & 2.21539 & 2.80568$\times 10^{-1}$ & 5.33909$\times 10^{-3}$\\
13 & 1.95528 & 1.83646 & 2.32824$\times 10^{-1}$ & 4.00920$\times 10^{-3}$\\
14 & 1.22395 & 1.14860 & 1.45535$\times 10^{-1}$ & 2.71298$\times 10^{-3}$\\
15 & 7.69355$\times 10^{-1}$ & 7.22687$\times 10^{-1}$ & 9.16398$\times 10^{-2}$ & 1.47368$\times 10^{-3}$\\
16 & 3.09857$\times 10^{-1}$ & 2.90757$\times 10^{-1}$ & 3.68732$\times 10^{-2}$ & 6.84761$\times 10^{-4}$\\
17 & 1.25355$\times 10^{-1}$ & 1.17772$\times 10^{-1}$ & 1.49401$\times 10^{-2}$ & 1.79301$\times 10^{-4}$\\
\hline\hline
\end{tabular}
\end{center}
\end{table}

\begin{table}[ht!]\caption{Chebyshev coefficients, $c_{np}$, for ionization in the STA $(1)^2(2)^6-(1)^2(2)^5$.}\label{Cheb-coef-EII-SC1-SC2}
\begin{center}
\begin{tabular}{c|c|c|c|c}
\hline\hline
$n\;\downarrow\;p \rightarrow$ & 0 & 1 & 2 & 3\\
\hline
0 & -1.04853$\times 10^{1}$ & -1.51446&4.67629$\times 10^{-2}$ & -5.62976$\times 10^{-2}$\\
1 & 5.60183$\times 10^{-1}$ & 8.57859$\times 10^{-1}$ & 1.50185$\times 10^{-1}$ & -6.81339$\times 10^{-2}$\\
2 & -4.53999$\times 10^{-1}$ & -2.96049$\times 10^{-1}$ & 1.06101$\times 10^{-4}$ & -6.15067$\times 10^{-2}$\\
3 & 2.01629$\times 10^{-1}$ & 2.73908$\times 10^{-1}$ & 6.51583$\times 10^{-2}$ & -4.82368$\times 10^{-2}$\\
4 & -1.28910$\times 10^{-1}$ & -5.52012$\times 10^{-2}$ & 1.19611$\times 10^{-2}$ & -3.16403$\times 10^{-2}$\\
5 & 5.06395$\times 10^{-2}$ & 7.38574$\times 10^{-2}$ & 1.83721$\times 10^{-2}$ & -1.36799$\times 10^{-2}$\\
6 & -2.67176$\times 10^{-2}$ & -3.48045$\times 10^{-2}$ & -7.54999$\times 10^{-3}$ & 4.79832$\times 10^{-3}$\\
7 & 2.47525$\times 10^{-2}$ & -1.75625$\times 10^{-2}$ & -1.46265$\times 10^{-2}$ & 2.07111$\times 10^{-2}$\\
8 & 8.29504$\times 10^{-3}$ & -5.78868$\times 10^{-2}$ & -2.76772$\times 10^{-2}$ & 3.31056$\times 10^{-2}$\\
9 & 2.32418$\times 10^{-2}$ & -5.78104$\times 10^{-2}$ & -3.19893$\times 10^{-2}$ & 4.03668$\times 10^{-2}$\\
10 & 1.81621$\times 10^{-2}$ & -6.57739$\times 10^{-2}$ & -3.39940$\times 10^{-2}$ & 4.18381$\times 10^{-2}$\\
11 & 1.95842$\times 10^{-2}$ & -5.80930$\times 10^{-2}$ & -3.13781$\times 10^{-2}$ & 3.88802$\times 10^{-2}$\\
12 & 1.46496$\times 10^{-2}$ & -4.81492$\times 10^{-2}$ & -2.53894$\times 10^{-2}$ & 3.13121$\times 10^{-2}$\\
13 & 1.13777$\times 10^{-2}$ & -3.47704$\times 10^{-2}$ & -1.91148$\times 10^{-2}$ & 2.32361$\times 10^{-2}$\\
14 & 6.68679$\times 10^{-3}$ & -2.12808$\times 10^{-2}$ & -1.14603$\times 10^{-2}$ & 1.40156$\times 10^{-2}$\\
15 & 3.87442$\times 10^{-3}$ & -1.15774$\times 10^{-2}$ & -6.78132$\times 10^{-3}$ & 7.93885$\times 10^{-3}$\\
16 & 1.40267$\times 10^{-3}$ & -4.32279$\times 10^{-3}$ & -2.45144$\times 10^{-3}$ & 2.89560$\times 10^{-3}$\\
17 & 5.55723$\times 10^{-4}$ & -1.44421$\times 10^{-3}$ & -1.05049$\times 10^{-3}$ & 1.09235$\times 10^{-3}$\\
\hline\hline
\end{tabular}
\end{center}
\end{table}

Let us focus on the STA $(1)^2(2)^6-(1)^2(2)^5(3)^1$, where each superconfiguration encompasses several configurations (see Eqs. (\ref{eq:SC})-(\ref{eq:SC'})). This STA involves more than 2,100 transitions between relativistic energy levels. As in the previous cases describing excitation, the temperature and ion charge values lie within the ranges 50$-$3000 eV and 6$-$28, respectively. The rate coefficient $q_{\Xi_I-\Xi_F}$ is given by Eqs. (\ref{eq:q_xi-xip}) or (\ref{eq:q_xi-xi2p}). It is calculated from the cross sections provided by the FAC code. The fit of the calculated rates utilizes expansions of Chebyshev polynomials in two dimensions, with degrees of 17 for temperature and 3 for ion charge.

\begin{figure}[ht!]
    \centering
    \includegraphics[scale=.5]{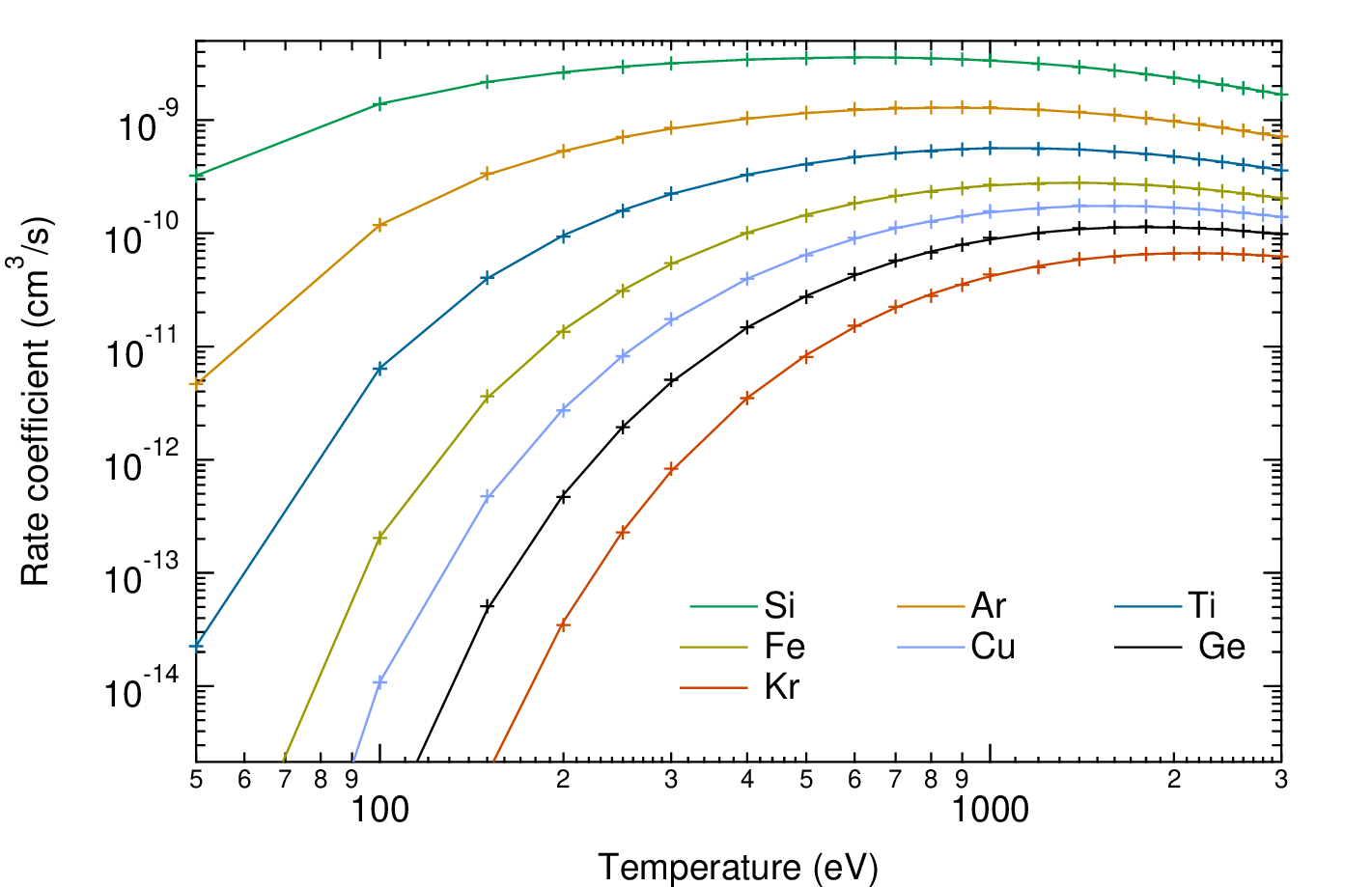}
    \caption{Excitation rate coefficient in the STA $(1)^2(2)^6-(1)^2(2)^5(3)^1$ in O-like ions. Solid line: direct calculations using Eq. (\ref{eq:q_xi-xip}), cross markers: fit using Chebyshev polynomials.}\label{EIE_SC1-SC2_ext}
\end{figure}

Figure \ref{EIE_SC1-SC2_ext} represents the temperature dependence of the excitation rate in the above STA. The Chebyshev coefficients $c_{np}$ are provided in Table \ref{Cheb-coef-EIE-SC1-SC2}. The agreement between the fit and the calculated rates is satisfactory at all temperatures and for all ions. Comparison with Fig. \ref{EIE_2p-3s_extended} shows that the STA rates are around one order of magnitude larger than for the transition array $2p^4-2p^33s^1$.

\begin{figure}
    \centering
    \includegraphics[scale=.5]{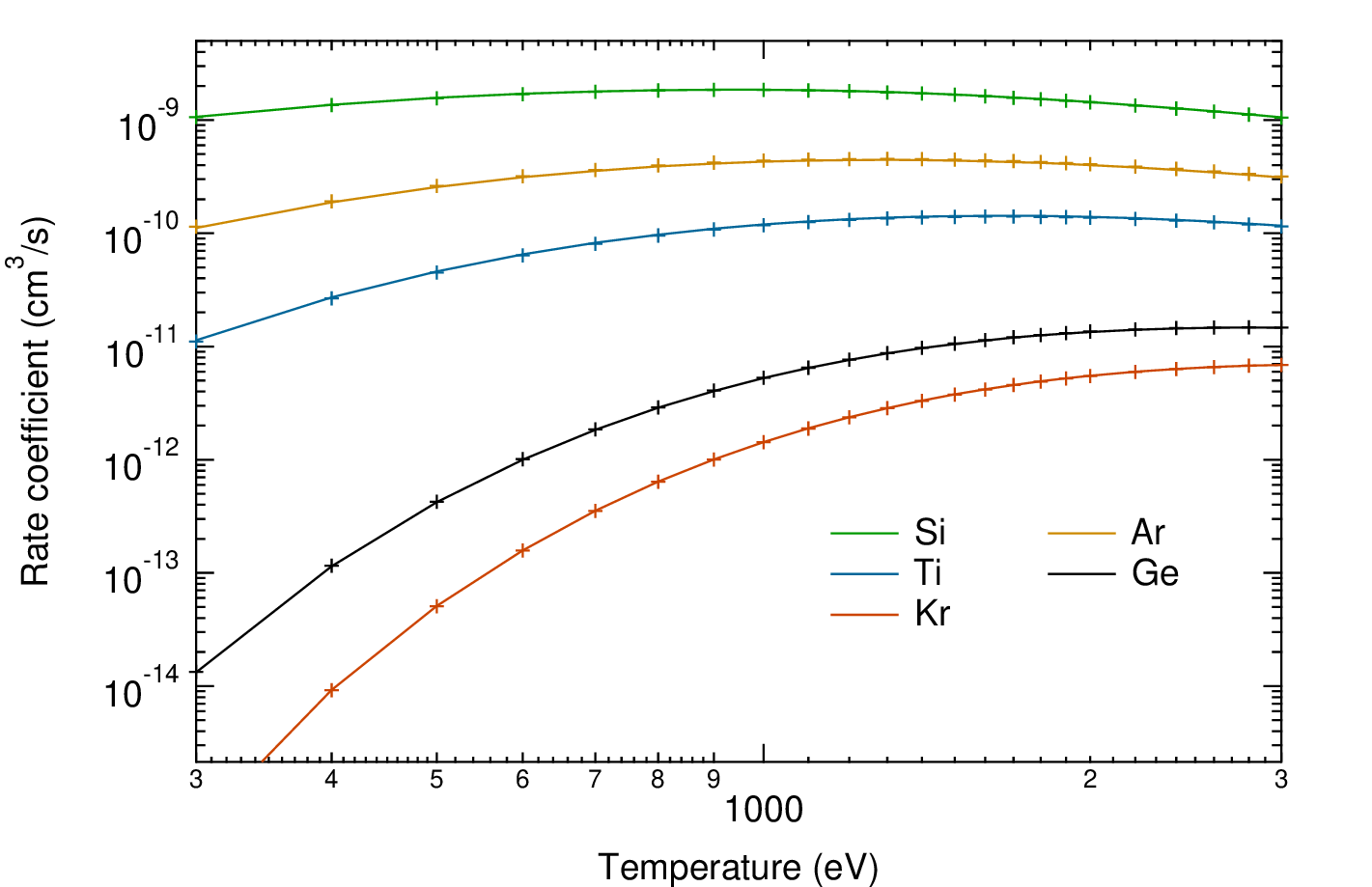}
    \caption{Ionization rate coefficient in the STA $(1)^2(2)^6-(1)^2(2)^5$, in O-like ions. Solid line: direct calculations using Eq. (\ref{eq:q_xi-xip}), cross markers: fit using Chebyshev polynomials.}\label{EII_SC1-SC2}
\end{figure}

Figure \ref{EII_SC1-SC2} illustrates the temperature dependence of the ionization rate for the STA $(1)^2(2)^6-(1)^2(2)^5$. The Chebyshev coefficients $c_{np}$ are listed in Table \ref{Cheb-coef-EII-SC1-SC2}. The fit shows strong agreement with the calculated rates across all investigated temperatures and ions. Comparison with Figs. \ref{EII_2p4-2p3} and \ref{EII_2p4-2s12p4}, corresponding to the transition arrays $2p^4-2p^3$ and $2p^4-2s^12p^4$, respectively, reveals that the contribution of the former is approximately 74 \% of that of the STA, while the latter contributes about 22 \%. The contribution of the other transition arrays is much smaller ($\simeq$ 4 \%).

In Section \ref{Clenshaw algorithm}, we present an alternative to the expansions in terms of Chebyshev polynomials, allowing for fast interpolation with respect to temperature and ion charge. 

\section{The Clenshaw algorithm}\label{Clenshaw algorithm}

The Clenshaw algorithm provides a powerful tool to evaluate two-dimensional expansions in terms of Chebyshev polynomials \cite{Clenshaw1955,Clenshaw1957}. It is somewhat faster than a barycentric interpolation as it requires only additions and multiplications, and can also be vectorized quite easily.

\subsubsection{One-dimensional Clenshaw algorithm}

We first consider a discrete function $f$ of the variable $x$, known for a set of $x$ values only. The idea is to approximate $f$ by an analytical form, which consists in an expansion in terms of Chebyshev polynomials (see. Eq. (\ref{eq:1D-expansion}). The approximation yields Chebyshev coefficients $c_n$. In the following, we show that the analytical form of $f$ can also be obtained by using a recursive method. In fact, knowing the Chebyshev coefficients, we can compute $f(x)$ by using a single loop to calculate the following functions $b_n(x)$, which are defined recursively \cite{Ledoux2020}:
\begin{equation*}
b_n(x) =	
\begin{cases}
0 & \text{ if } n=N+2 \\
0 & \text{ if } n=N+1 \\
2xb_{n+1}(x)-b_{n+2}(x)+c_n & \text{ if } 1\leq n\leq N \\
2xb_{1}(x)-b_{2}(x)+2c_0 & \text{ if } n=0. \\ 
\end{cases}
\end{equation*} 
and finally one gets
\begin{equation*}
    f(x) = \frac{1}{2}\left[b_0(x)-b_2(x)\right].
\end{equation*}
This approach is numerically stable and avoids directly computing powers of $x$, making it an efficient method for evaluating Chebyshev expansions. With the aim of investigating the variation of the rate coefficient with temperature and ion charge, we extend the algorithm to two dimensions.

\subsubsection{Two-dimensional Clenshaw algorithm}

In the following, we give an analytical expression of the rate coefficient as a function of temperature and ion charge. To this end, we investigate a powerful alternative to the conventional expansion in terms of products of two Chebyshev polynomials. Given the Chebyshev coefficients $c_{np}$, the alternative method extends the Clenshaw algorithm to two dimensions, which requires the iterative application of the algorithm in both dimensions to the function $f(\bar{T}, \bar{z})$:
\begin{equation*}
f(\bar{T}, \bar{z}) = \sum_{n=0}^{N} \sum_{p=0}^{P} c_{np} T_n(\bar{T}) T_p(\bar{z}).
\end{equation*}
It turns out that the two-dimensional implementation is somewhat tricky, in the sense that it is not the superposition of two one-dimensional algorithms. The Clenshaw loops for variables $\bar{T}$ and $\bar{z}$ are actually nested.

Knowing the Chebyshev coefficients $c_{np}$, the evaluation proceeds as follows. We first set a value for $\bar{z}$ and, for a given $p$ value, we perform a recurrence over $n$. For a given normalized temperature $\bar{T}$, we compute $a_n(\bar{T}|\bar{z}$) for $n=N,\ N-1,\cdots, 1$:
\begin{equation*}
    a_{N+2}(\bar{T}|\bar{z}) = 0, \quad a_{N+1}(\bar{T}|\bar{z}) = 0, \quad a_0(\bar{T}|\bar{z})=2\bar{T}a_{1}(\bar{T}|\bar{z})-a_{2}(\bar{T}|\bar{z})+2c_{0p},
\end{equation*}
\begin{equation*}
    a_n(\bar{T}|\bar{z}) = 2\bar{T} a_{n+1}(\bar{T}|\bar{z}) - a_{n+2}(\bar{T}|\bar{z})+c_{np} \quad \text{for} \quad n = N, N-1, \dots, 1,
\end{equation*}
where $(A|B)$ means that we vary $A$, through the recurrence, with fixed $B$.
We then obtain the intermediate function $g_P$, representing the variation with the net ion charge $z$:
\begin{equation*}
    g_p(\bar{T}|\bar{z})=\frac{1}{2}\left[a_0(\bar{T}|\bar{z})-a_2(\bar{T}|\bar{z})\right].
\end{equation*}
Knowing $g_p(\bar{T}|\bar{z})$, we perform a recurrence on $p$ by computing intermediate values $b_p(\bar{z}|\bar{T})$ for
$p=P,\ P-1,\cdots, 1$:
\begin{equation*}
    b_{P+1}(\bar{z}|\bar{T}) = 0, \quad b_{P+2}(\bar{z}|\bar{T})= 0, \quad b_0(\bar{z}|\bar{T})=2\bar{z}b_{1}(\bar{z}|\bar{T})-b_{2}(\bar{z}|\bar{T})+2g_0(\bar{T}|\bar{z}),
\end{equation*}
\begin{equation*}
    b_p(\bar{z}|\bar{T}) = 2\bar{z} b_{p+1}(\bar{z}|\bar{T}) - b_{p+2}(\bar{z}|\bar{T})+g_{p}(\bar{T}|\bar{z}) \quad \text{for} \quad p = P, P-1, \dots, 1.
\end{equation*}
and finally
\begin{equation*}
    f(\bar{T},\bar{z})=\frac{1}{2}\left[b_0(\bar{z}|\bar{T})-b_2(\bar{z}|\bar{T})\right].
\end{equation*}
It is clear that the two sets, $\lbrace{a_n}\rbrace$ and $\lbrace{b_p}\rbrace$, are expressed in terms of the Chebyshev coefficients $c_{np}$. The Clenshaw algorithm is efficient because it reduces the complexity from $O(N^2) $ (for a direct sum) to $O(N) $ for each dimension. In two dimensions, this allows us to evaluate the bivariate series in $O(NP) $ for a grid of size $(N+1) \times (P+1)$. 

\subsubsection{Application}
We use the Clenshaw algorithm to calculate rates for an ion that was not included in the Chebyshev polynomial fitting process. Specifically, we focus on Ca$^{12+}$, whose net charge, $z=12$, lies within the $z$-interval used in the fitting. Temperatures in the range of 50$-$3000 eV are selected, excluding, of course, those already considered in the fitting.

\begin{figure}[ht!]
    \centering
    \includegraphics[scale=.5]{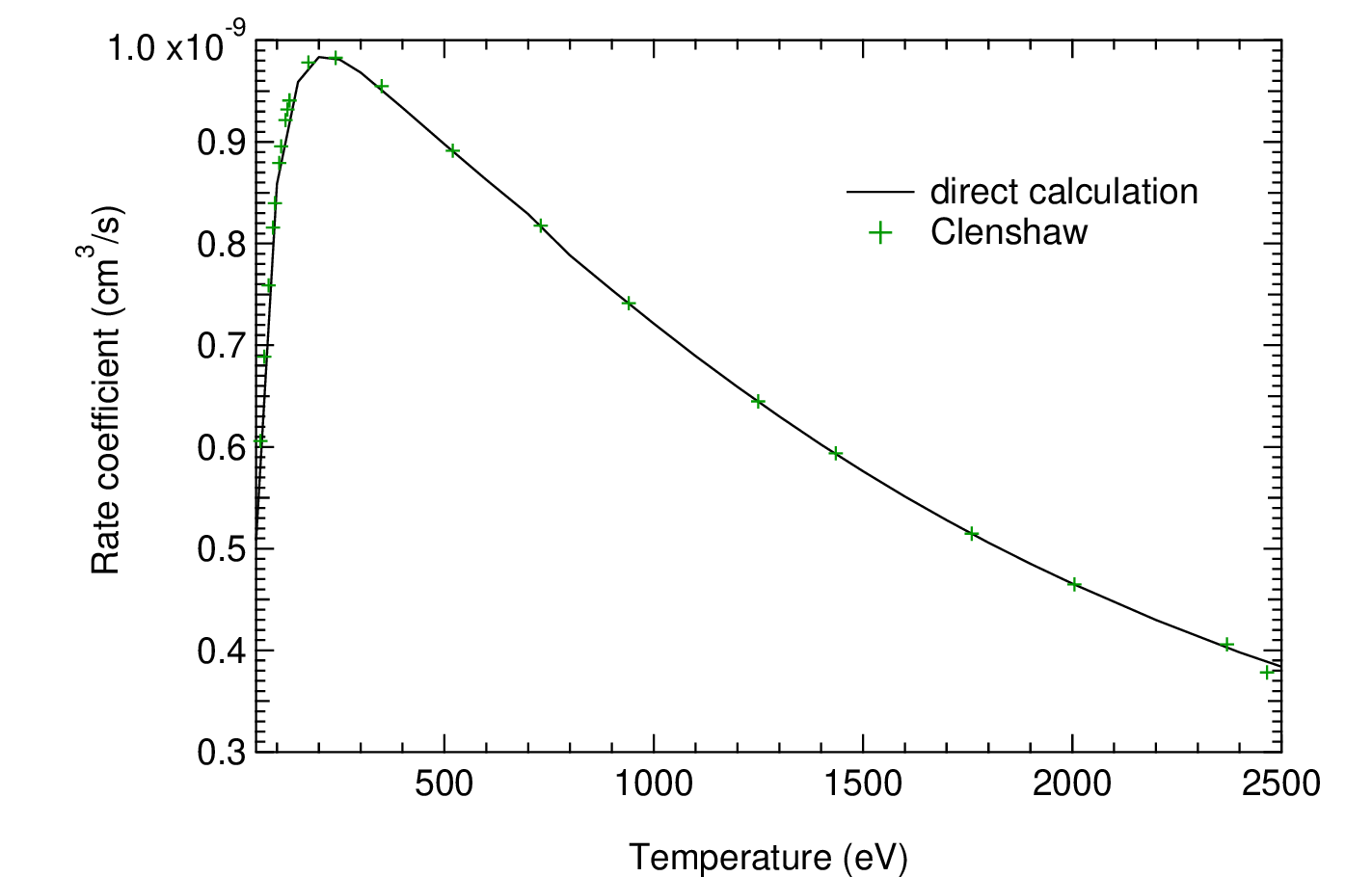}
    \caption{Excitation rate coefficient in the $2p^4-2s^1 2p^5$ transition array in O-like calcium. Solid line: direct calculations using Eq. (\ref{eq:q_CC'}), cross markers: calculations using the Clenshaw algorithm.}\label{EIE_Ca_2s-2p_ext}
\end{figure}

Consider the $2p^4-2s^{1}2p^5$ transition array. In Fig. \ref{EIE_Ca_2s-2p_ext}, we present the variation of the EIE rate with temperature. The solid line shows the rate calculated from the cross sections provided by FAC (direct calculations, from Eq. (\ref{eq:q_CC'})). The cross markers represent the rate calculated using the Clenshaw algorithm for selected temperatures. As can be seen, there is good agreement between the two approaches, providing confidence in both the Chebyshev polynomial fit and the Clenshaw algorithm.

Using the coefficients $c_{np}$, the Clenshaw algorithm enables efficient computation of EII and EIE rates for any temperature and any O-like ion charge within the $z$ interval of 6–28. This method is significantly faster than direct calculations, approximately by a factor of 2, making it more suitable for generating large datasets, which are essential for collisional-radiative modeling in non-LTE plasmas. Although not encountered in our study, the Clenshaw algorithm helps avoid numerical issues that can arise from the alternating summation of high-degree Chebyshev polynomial monomials.

We confirm that the Clenshaw algorithm provides rates identical to those derived from Chebyshev polynomial expansions, across all temperatures and ion charges within the fitting intervals.

\section{Conclusion and prospective}
We have calculated excitation and ionization rates for transition arrays and super-transition arrays over a broad range of temperatures for several O-like ions, some of which are relevant in the context of hot and dense plasmas. We investigated cases involving large sets of transitions. Consequently, direct calculations using atomic codes such as FAC \cite{Gu2008}, HULLAC \cite{Barshalom2001}, \emph{etc.}, require substantial computational effort. Due to the complexity of fine-structure and radiative-collisional calculations, many computational codes use statistical methods, grouping levels into configurations or even superconfigurations, and defining rates between these entities. In order to obtain the rate for any temperature and/or ion charge in definite intervals, we have fitted the direct calculations with two-dimensional expansions in terms of Chebyshev polynomials. The obtained Chebyshev coefficients allow us to generalize the Clenshaw algorithm in order to address two-dimensional issues. The Clenshaw algorithm significantly reduces computation time while maintaining accurate rate determination for any temperature and/or ion charge within the range covered by the two-dimensional Chebyshev fitting procedure. As expected, these rates are identical to those obtained with the two-dimensional Chebyshev polynomial fit. Unfortunately, the Chebyshev polynomial expansion is not reliable for extrapolation beyond the $z$ interval. For example, in O-like xenon ($Z=54$ and $z=46$), the EIE rates for the $2s–2p$ transition array deviate from direct calculations by up to 23 \%. We are currently investigating more advanced methods to achieve the same level of accuracy as obtained in interpolation.\\ 
\indent When studying hot and dense plasmas, the excitation-autoionization process should be considered along with direct ionization \cite{Qi2002}. However, preliminary calculations showed that, for O-like ions, the rate of this process is negligible compared to direct ionization due to the small excitation rate to doubly excited levels. It is also important to examine the impact of non-Maxwellian electron distributions and density effects. These issues will be addressed in future work.\\
\indent The authors can provide, upon request, a code (in {\sc Python} or {\sc Mathematica} \cite{Mathematica} programming languages) for calculating excitation or direct ionization rates, either in transition arrays or super-transition arrays.

\end{document}